\documentclass[twocolumn]{aastex631}

\usepackage{CJKutf8}
\begin{document}
\begin{CJK}{UTF8}{gbsn}

\title{Radial Wave in the Galactic Disk: New Clues to Discriminate Different Perturbations}

\author[0000-0001-6655-854X]{Chengye Cao}
\affiliation{Department of Astronomy,School of Physics and Astronomy, Shanghai Jiaotong University,\\
800 Dongchuan Road, Shanghai 200240, China}

\author[0000-0001-5017-7021]{Zhao-Yu Li}
\affiliation{Department of Astronomy,School of Physics and Astronomy, Shanghai Jiaotong University,\\
800 Dongchuan Road, Shanghai 200240, China}
\affiliation{Key Laboratory for Particle Astrophysics and Cosmology(MOE)/Shanghai Key Laboratory for Particle Physics and Cosmology, \\ Shanghai 200240,Peopleʼs Republic of China}
\email{lizy.astro@sjtu.edu.cn}

\author[0000-0002-4236-3091]{Ralph Sch{\"o}nrich}
\affiliation{Mullard Space Science Laboratory, University College London, Holmbury St Mary, Dorking, Surrey RH5 6NT, UK}

\author[0000-0003-2595-5148]{Teresa Antoja}
\affiliation{Departament de Física Quàntica i Astrofísica (FQA), Universitat de Barcelona (UB), c. Martí i Franquès, 1, 08028, Barcelona, Spain}
\affiliation{Institut de Ciències del Cosmos (ICCUB), Universitat de Barcelona (UB), c. Martí i Franquès, 1, 08028, Barcelona, Spain}
\affiliation{Institut d'Estudis Espacials de Catalunya (IEEC), Edifici RDIT, Campus UPC, 08860 Castelldefels (Barcelona), Spain}



\begin{abstract}
Decoding the key dynamical processes that shape the Galactic disk structure is crucial for reconstructing the Milky Way's evolution history. The second Gaia data release unveils a novel wave pattern in the $L_Z-\langle V_R\rangle$ space, but its formation mechanism remains elusive due to the intricate nature of involved perturbations and the challenges in disentangling their effects. Utilizing the latest Gaia DR3 data, we find that the $L_Z-\langle V_R\rangle$ wave systematically shifts toward lower $L_Z$ for dynamically hotter stars with larger $J_Z$ values. The amplitude of this phase shift between stars of different dynamical hotness ($\Delta L_Z$) peaks at around $\mathrm{2100\,km\,s^{-1}\,kpc}$. To differentiate the role of different perturbations, we perform three sets of test particle simulations, wherein a satellite galaxy, transient spiral arms, and a bar plus the transient spiral arms act as the sole perturber, respectively. Under the satellite impact, the phase shift amplitude $\Delta L_Z$ decreases toward higher $L_Z$, which we interpret through a toy model of radial phase mixing. While neither the transient spiral arms nor the bar generates an azimuthally universal phase shift variation pattern, combining the bar and spirals generates a characteristic $\Delta L_Z$ peak at the 2:1 Outer Lindblad Resonance (OLR) of the bar, qualitatively resembling the observed feature. Therefore, the $L_Z-\langle V_R\rangle$ wave is more likely of internal origin. Furthermore, linking the $\Delta L_Z$ peak to the 2:1 OLR offers a novel approach to constraining the pattern speed of the Galactic bar, supporting the long/slow bar model.

\end{abstract}



\section{Introduction} \label{sec: intro}
The accurate astrometric information provided by the Gaia mission has revolutionized the field of Milky Way dynamics. Three-dimensional positions and velocities of millions of stars provided by synergies between Gaia DR2 \citep{gaia_dr2_18} and large spectroscopic surveys have led to the discovery of a series of phase space substructures in the Galactic disk, including the diagonal ridges in the $R-V_{\phi}$ plane \citep{antoja_etal_18,kawata_etal_18,ramos_etal_18}. The potential contributing perturbers for generating these $R-V_{\phi}$ ridges and associated velocity substructures include the Galactic bar \citep{muhlbauer_03,Chakrabarty_07,antoja_etal_18,monari_etal_19,fragkoudi_etal_19}, spiral arms \citep{hunt_etal_18,hunt_etal_19,quillen_etal_18,khanna_etal_19}, and the Sagittarius-like satellite \citep[Sgr;][]{minchev_etal_09,gomez_etal_12,khanna_etal_19,laport_etal_19}. Understanding the physical origin of these ridges is crucial for reconstructing the Milky Way's dynamical evolution history.

However, a consensus on the formation mechanisms of these ridge-like structures remains elusive, owing to the complexities associated with various perturbations and the challenges involved in disentangling their effects. Test particle simulations conducted by \cite{hunt_etal_19} have demonstrated that the combination of a bar with arbitrary pattern speeds and transient spirals can qualitatively reproduce the observed $R-V_\phi$ ridges, thereby making it exceedingly difficult to isolate the individual effects of these perturbers. According to the $N$-body simulation sets of \cite{khanna_etal_19}, both transient spiral arms excited in isolation and a satellite perturber at the mass scale of $10^{10}M_\odot$ generate ridges qualitatively similar to the observations in the density and $\langle V_R \rangle$ map. This degeneracy necessitates the identification of new discriminating features.

The ridge-like structures exhibit a strong connection with radial motion, as evidenced by the mean radial velocity $\langle V_R \rangle$ map in the $R-V_{\phi}$ plane \citep{fragkoudi_etal_19,hunt_etal_19,khanna_etal_19,wang_etal_20}. Notably, these structures align approximately along the lines of constant angular momentum $(L_Z=R\times V_{\phi})$, mirroring the morphology of ridges in the number density map. Utilizing orbit integration in an $N$-body potential, \cite{fragkoudi_etal_19} demonstrated that a long $R-V_{\phi}$ ridge, accompanied by a change in the $\langle V_R \rangle$ direction, could pinpoint the location of the 2:1 outer Lindblad resonance (hereafter OLR) of the bar, reinforcing the physical connection between ridges in these two projections. Consequently, the $\langle V_R \rangle$ corrugation binned in angular momentum $L_Z$ can be viewed as the one-dimensional projection of the 2D $R-V_{\phi}$ ridges, simplifying comparative analysis while retaining the essential physical information.

First discovered by \cite{friske_19}, this $L_Z-\langle V_R\rangle$ wave displays systematic displacement toward lower $L_Z$ for stars with higher vertical energy $(E_Z)$, suggesting a phase shift among the wave pattern of stars with different dynamical hotness. \cite{friske_19} attributed this feature to orbital resonances, which raises an interesting question on the existence of such phase shift when subject to other perturbations like the satellite pericenter passage. Moreover, it remains unknown whether the variation in the phase shift amplitude with $L_Z$ encodes information about its origin.

Previous works have shown that phase space substructures could vary with dynamical hotness. In the $\langle V_R\rangle$ color coded $R-V_{\phi}$ space, \cite{wang_etal_20} found that some ridges vary among populations of different stellar ages whereas others do not. Since stellar age (or metallicity) is a crude proxy of the dynamical hotness, interpreting this dichotomy from the dynamical perspective could better uncover its physical origin. In the vertical direction, \cite{li_shen_20} found the $Z-V_Z$ phase spiral becomes less prominent or even absent for those stars with higher $J_R$ (dynamically hotter in the radial direction). Analogously, an analytic model of \cite{laporte_etal_20} also demonstrated that the $R-V_{\phi}$ ridges generated by bar resonances are more prominent for the dynamically colder population. 

To explore the connection between the $R-V_{\phi}$ ridges and spiral arms arising from a single satellite impact, \cite{ramos_etal_22} developed an analytical model and found that the $R-V_\phi$ ridges exhibit a $V$-shaped morphology in both test particle and $N$-body simulations. However, an intuitive explanation for the formation of such morphology remains lacking. Furthermore, an undulating $L_Z-\langle V_R\rangle$ wave emerges simultaneously, with its frequency increasing during the phase mixing. Fourier transform of the $L_Z-\langle V_R\rangle$ wave reveals two frequency peaks, attributed to perturbations occurring less than 0.4 Gyr ago and $0.7-1.8$ Gyr ago, respectively. However, their conclusions only hold if the satellite is the sole perturber. Both \cite{antoja_etal_18} and \cite{khanna_etal_19} used toy models of winding spirals to mimic the $R-V_\phi$ ridges and found qualitative agreement with the observed $R-V_\phi$ density map, without accounting for its correlation with the $\langle V_R\rangle$ map. A deeper understanding of the role of other perturbing mechanisms in generating the $L_Z-\langle V_R \rangle$ wave is still lacking.

We present a novel perspective to ascertain the origin of the $L_Z-\langle V_R \rangle$ wave by analyzing its dependence on dynamical hotness. We quantify this dependence through the phase shift between waves of varying dynamical hotness. After presenting the observational phase shift variation trend with $L_Z$ in Section \ref{subsec:data}, we set up two groups of test particle simulations in Section \ref{subsec:tps} to differentiate the role of internal and external perturbers in generating the phase shift variation trend. We extract the key features of the simulated phase shift variation pattern in Sections \ref{subsec:ext} and \ref{subsec:int}, accompanied by our physical interpretation, including a toy model of radial phase mixing for the case of external perturbation. We discuss the caveats and future work in Section \ref{sec:discu}, and summarize our main findings in Section \ref{sec:conclu}.

\section{$L_Z-\langle V_R\rangle$ wave in Gaia DR3} \label{subsec:data}

Among 33 million stars with astrometry and line-of-sight velocity measurement \citep{gdr3_rv} from Gaia DR3 \citep{gdr3}, we obtain 26,611,026 sources meeting the criteria for reliable \texttt{StarHorse} \citep{anders_etal_22} distances, with $\texttt{fidelity} >0.5$ \citep{rybizki_etal_22} and \texttt{sh\_outflag}=``0000". \texttt{StarHorse} is a Bayesian code that leverages astrometric, photometric, and spectroscopic data from multiple surveys to derive the cumulative distribution function of astrophysical parameters, including the Heliocentric distance (whose median is \texttt{dist50} column). We adopt a distance of 8.275 kpc between the Sun and the Galactic center \citep{gravity_21}, with the Sun situated 20.8 pc above the Galactic midplane \citep{bennett_19}. For the motion of Sgr A*,  we use a radial velocity of $\mathrm{-8.4\,km\,s^{-1}}$ \citep{gravity_21} and a proper motion in the International Celestial Reference System (ICRS) frame $\mathrm{\mu_{ICRS}=(-3.16,-5.59)\,mas\,yr^{-1}}$ \citep{reid_20}. Combining these measurements yields the total solar velocity with respect to the Galactic center $v_{\odot}=(8.7,251.5,8.4)\, \mathrm{km\, s^{-1}}$ \citep{hunt_etal_22}. Our cut in the azimuthal range $|\phi-\phi_{\odot}|<0.2 \, \mathrm{rad}$ gives a sample size of 19,279,240. Using the “St$\ddot{a}$ckel Fudge” method \citep{binney_12} incorporated in the \texttt{agama} package \citep{agama_19}, we calculate the action-angle-frequency quantities for the entire sample, employing \texttt{MWPotential2014} \citep{galpy_15} as the Milky Way potential model. We choose not to consider the selection function effect in our study due to its minor impact on the mean velocity map. 

\begin{figure}[ht]
    \centering
    \includegraphics[scale=0.5]{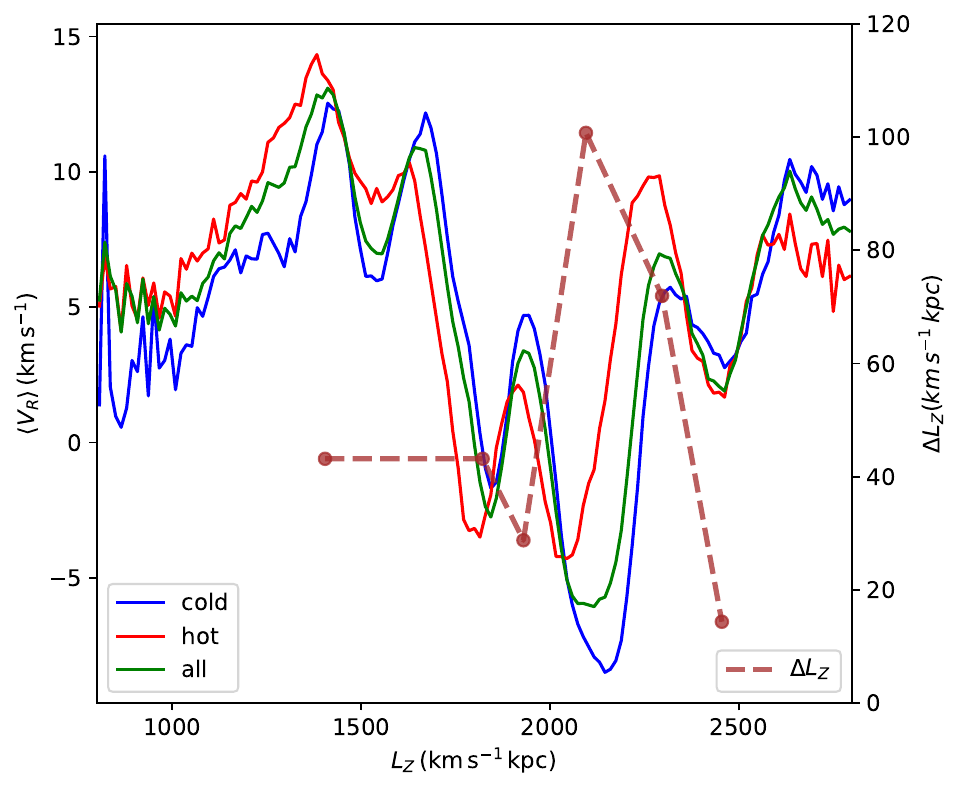}
    \caption{Shape and phase shift between particles of the $L_Z-\langle V_R\rangle$ wave with different orbit hotness of the $L_Z-\langle V_R\rangle$ wave in the Gaia DR3 data. The green line is the $L_Z-\langle V_R\rangle$ wave of the whole sample. The $L_Z-\langle V_R\rangle$ wave patterns composed of dynamically cold ($J_Z\mathrm{<3 \thinspace km \thinspace s^{-1}\thinspace kpc}$) and hot ($J_Z\mathrm{>12 \thinspace km \thinspace s^{-1}\thinspace kpc}$) subpopulations are depicted in red and blue lines, respectively. The brown dashed line displays the phase shift amplitude ($\Delta L_Z$, shown in the right axis) variation pattern, which peaks at around 2100 $\mathrm{km\,s^{-1}\,kpc}$. }
    \label{fig:f_peak}
\end{figure}

As depicted by the solid green line in Figure \ref{fig:f_peak}, the overall shape of the $L_Z-\langle V_R \rangle$ wave for the whole sample is consistent with the previous work of \cite{ramos_etal_22}. It is also consistent with the wave shape of \cite{friske_19} after accounting for the opposite definition of positive $V_R$. To examine the morphological variation of the $L_Z-\langle V_R\rangle$ wave with dynamical hotness, we categorize stars based on their vertical action $J_Z$. It is a better-conserved quantity than vertical orbit energy $E_Z$, as pointed out by \cite{friske_19}. We do not adopt another adiabatic invariant $J_R$ due to its tight correlation with the planar phase space coordinates, which could introduce bias in the resulting $L_Z-\langle V_R\rangle$ signal.

In our sample division based on dynamical hotness, stars with $J_Z < 3\, \mathrm{km\,s^{-1}\,kpc}$ are considered dynamically ``cold," while those with $J_Z > 12\, \mathrm{km\,s^{-1}\,kpc}$ are considered dynamically ``hot". This classification guarantees a sufficient difference in dynamical behavior between the two subsamples, allowing for measurable phase shifts in their wave patterns, while also ensuring that the morphological difference between the wave patterns is not substantial enough to introduce systematic bias in the phase shift measurement. The investigated $L_Z$ range is limited to $\mathrm{[800,2800]\,km\,s^{-1}\,kpc}$, beyond which the observational uncertainty is too large for accurate phase shift measurement. The extrema points of the ``hot" wave systematically shift toward lower $L_Z$ compared to the ``cold" wave, qualitatively consistent with the results of \cite{friske_19} who used vertical energy $E_Z$ as a proxy for dynamical hotness. Both the ``hot" and ``cold" waves exhibit similar shapes, except at the high $L_Z$ end near $\mathrm{2800\, km \, s^{-1}\, kpc}$. Due to the larger measurement uncertainty of the data, whether the two waves are truly phase aligned at the high $L_Z$ end is inconclusive. 

We define the phase shift amplitude as the difference between the locations of local extrema points, $\Delta L_Z=L_{Z,cold}-L_{Z,hot}$. The subscripts correspond to the dynamically cold and hot subsamples of the whole distribution, which is dissected based on the values of $J_Z$. To analyze the variation trend of $\Delta L_Z$ with $L_Z$ (referred to as $\Delta L_Z$ variation trend for brevity in the following text), we define the mean $L_Z$ location of the extrema as $\bar{L_Z}=(L_{Z,cold}+L_{Z,hot})/2$. Strictly speaking, quantifying the actual phase shift requires dividing $\Delta L_Z$ by the characteristic wavelength of the wave. Nevertheless, we adopt the value of $\Delta L_Z$ as a proxy for phase shift because the ever-changing shape of the wave across different $L_Z$ ranges severely complicates the task of extracting characteristic wavelengths and may introduce additional systematic bias.  First, we smooth the curve with a Gaussian kernel to mitigate the effect of small-scale noise. Visual inspection of the smoothed wave signal ensures that this process does not generate pseudo-oscillations that may compromise the $\Delta L_Z$ measurement. Then, we adjust the parameters of the \texttt{scipy.signal.find\_peaks} function (i.e. \texttt{distance}, \texttt{width} etc.) to find the suitable parameter set capable of identifying all noticeable extrema points of the $L_Z-\langle V_R\rangle$ wave pattern, and apply the same settings to all analyses with simulations. We discard those extrema points that have no counterparts in the $L_Z-\langle V_R$ wave of the other subpopulation. Furthermore, we set the \texttt{distance} parameter at 11 to avoid adjacent extrema points being too close to each other. The upper bound of the \texttt{width} parameter is also set at a value high enough to identify extrema points at higher $L_Z$ where the wavelength becomes longer in that range. We only include an extrema point when its \texttt{prominence} parameter is greater than 1 to mitigate the effect of small amplitude oscillation.

After smoothing the $L_Z-\langle V_R\rangle$ wave signal with the Gaussian kernel $(\sigma=2)$, we obtain the $\Delta L_Z$ variation trend illustrated by the brown dashed line in Figure \ref{fig:f_peak} (corresponding to the right axis). The phase shift amplitude ($\Delta L_Z$) is largest at the extrema point near $L_Z\sim 2100\,\mathrm{km\,s^{-1}\,kpc}$, which presents a prominent peak in the $\Delta L_Z$ variation trend. Using vertical orbit energy $E_Z$ as a proxy for the dynamical hotness, \cite{friske_19} also found the largest phase shift at around $L_Z\sim\,\mathrm{2100\,km\,s^{-1}\,kpc}$, which is qualitatively consistent with our results.

The phase shift $\Delta L_Z$ variation trend in observations raises some interesting questions: does this trend differ among different perturbations? If it does, can it be used to constrain the dynamical evolution history of the Milky Way disk? Can we find a way to qualitatively understand the physical origin of this trend? We will explore these questions in the following sections using test particle simulations.

\section{Test particle simulations} 
\subsection{Simulation Setup} \label{subsec:tps}
We now turn to test particle simulations involving internal and external perturbations separately, to examine the difference in the resulting phase shift $\Delta L_Z$ variation pattern of the $L_Z-\langle V_R\rangle$ wave. Compared to $N$-body simulations, test particle simulations neglect self-gravity, but do allow us to run more particles and test the effect of varying specific parameters on the orbit evolution.

\begin{figure}
    \centering
    \includegraphics[scale=0.5]{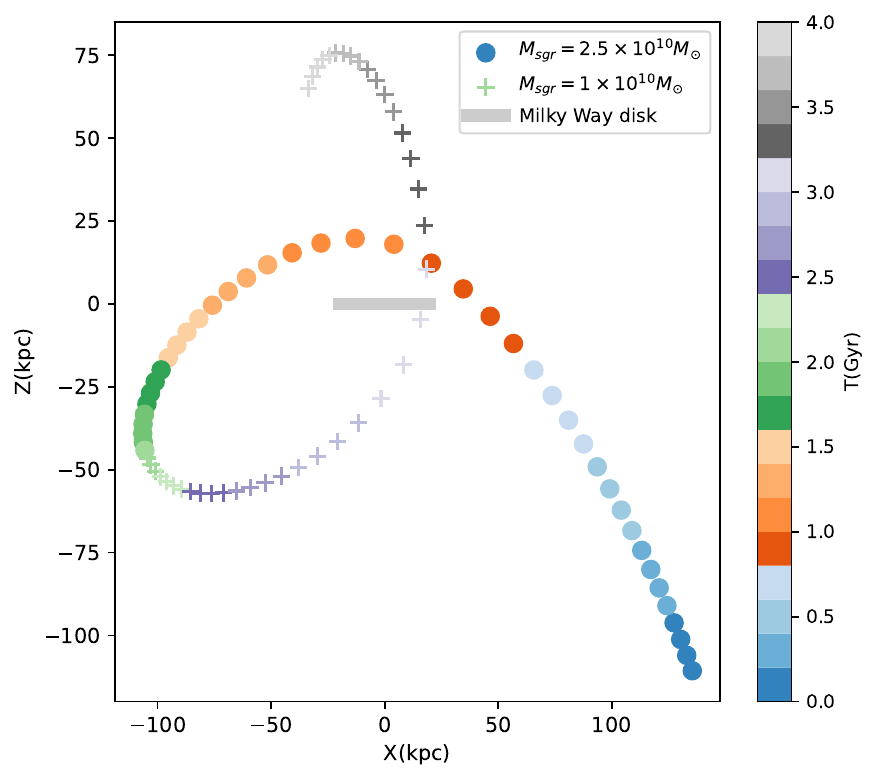}
    \caption{The orbit of the Sgr-like satellite in our test particle simulations, following the prescription in Section \ref{subsec:tps}. The circles and crosses represent the orbits of the satellite before and after reducing its mass from $2.5\times 10^{10}M_\odot$ to $1\times 10^{10}M_\odot$. Each point is color coded with time at an interval of 50 Myr. The bold gray line represents the Milky Way disk.}
    \label{fig:set_up}
\end{figure}

We sample eight million disk particles from the \texttt{quasiisothermal df} \citep{carlberg_85,binney_10,bin_mcm_11} implemented in \texttt{agama} \citep{agama_19}. These particles are distributed using a radial scale length of 3 kpc within the Milky Way potential model, \texttt{MWPotential2014} \citep{galpy_15}. The central velocity dispersion  $\mathrm{(\sigma_R|_{R=0},\sigma_Z|_{R=0})=(90,110)\,km\, s^{-1}}$ gives a disk that is dynamically colder than the actual Milky Way disk. The scale lengths of the $\sigma_R$ and $\sigma_Z$ profiles are 6 and 7 kpc, respectively. The $L_Z-\langle V_R\rangle$ wave signal is more prominent, which simplifies the task of identifying extrema points for the $\Delta L_Z$ measurement. To investigate the impact of dynamical hotness on the $L_Z-\langle V_R\rangle$ wave morphology, we define those particles with $J_Z\mathrm{<8\,km\,s^{-1}\,kpc}$ as the dynamically ``cold" population, and those with $J_Z\mathrm{>23\,km\,s^{-1}\,kpc}$ as the dynamically ``hot" population. Under such division, the particle numbers of the cold and hot populations are roughly the same. Note that this criterion of orbital hotness is different from that used in the observation. For the observation, the criterion is determined empirically to enable sufficient extrema points and coherent $L_Z-\langle V_R\rangle$ wave patterns for phase shift ($\Delta L_Z$) measurement. This is mainly due to the differences in the distribution function between the 
observation and simulation, and also the sample selection bias and velocity measurement error in the observational data, which is not present in the simulation results. Since our main goal is not to quantitatively reproduce the observational $\Delta L_Z$ variation trend but to qualitatively understand the physical origin of the $\Delta L_Z$ feature, we choose different $J_Z$ bounds for the sample division in the observation and simulation.

The external perturber (Milky Way satellite) is a $2.5\times 10^{10} M_{\odot}$ Plummer sphere with a half-mass radius of 3 kpc. Its position $\vec{x}=(4,8,18)\,\mathrm{kpc}$ and velocity $\vec{v}=(-339,-44,76)\,\mathrm{km}\,\mathrm{s}^{-1}$ at the first pericenter passage are from the E1 model of \cite{dl_vega_etal_15}. Backward orbit integration for 1 Gyr in the \texttt{MWPotential2014} potential using \texttt{galpy}'s \texttt{ChandrasekharDynamicalFrictionForce} routine to account for dynamical friction provides the initial condition for our simulation. At 2 Gyr, we reduce the mass and half-mass radius of the satellite to $1\times 10^{10} M_{\odot}$ and 1 kpc, respectively, to mimic the mass loss after the first pericenter passage. The second pericenter passage occurs at 3.1 Gyr when the satellite crosses the disk plane at $R=15\,\mathrm{kpc}$ with $\mathrm{V_Z\approx 300\,km\,s^{-1}}$. We adopt this simplified setup of the satellite mass loss for better simulation efficiency. A qualitatively similar prescription has been commonly used in previous works to model the satellite perturbation on the disc \citep[e.g.,][]{xu_etal_20,bennett_21,gandhi_etal_22}. Our tests, assuming a gradual mass loss history, produce qualitatively similar $L_z-\langle V_R\rangle$ wave features and $\Delta L_Z$ variation trend. The total integration time is set to 4 Gyr to cover these two pericenter passages. The full satellite orbit trajectory color coded by time is illustrated in Figure \ref{fig:set_up}.

Using test particle simulations, we also investigate the influence of internal perturbations with the combination of a steadily rotating bar and two-armed transient spirals. The spiral arm potential reaches its maximal amplitude at 1200 Myr. To differentiate the role of different perturbers in generating the $\Delta L_Z$ variation trend, we complement this simulation with a test where the transient spirals act as the sole perturber. 

In this test, the spiral potential has a higher amplitude and reaches the maximal value at the start. The two-armed transient spirals follow the default model described in Section 2.2 of \cite{hunt_etal_18} which follows the potential form given by \cite{cox_02}. The winding and decay of the spiral arms are configured by the \texttt{CorotatingRotationWrapperPotential} and \texttt{GaussianAmplitudeWrapperPotential} routines of \texttt{galpy}. Under this setup, spiral arms have a lifetime of about 100 Myr and approximately corotate with stars in the investigated radial range. The morphological parameters of the \texttt{SpiralArmsPotential} model are $\mathrm{N=2,R_s=0.3,C_n=1,H=0.125,\theta_{sp}=12^{\circ}}$ ($N,\,R_s,\,C_n,\,H,\,\mathrm{\theta_{sp}}$ represent the number of spiral arms, radial scale length, sinusoidal potential profile, scale height, and pitch angle, respectively). The bar is modeled as the 3D generalization of the Dehnen bar potential \citep{dehnen_00,monari_etal_16} with the same pattern speed (40 $\mathrm{km\, s^{-1}\, {kpc}^{-1}}$, invariable with time) and bar length (4.5 kpc) as the ``fiducial" model presented in \cite{trick_etal_21}. 

The integration time is 2 Gyr for the case with a bar and 0.4 Gyr without a bar (spirals only). Particle orbits are integrated using the \texttt{galpy} code \citep{galpy_15}. In the first 1 Gyr, the bar is the main driver of the phase space structure formation. Therefore, it can be used to illustrate the $\Delta L_Z$ variation due to the effect of the bar. A fully grown bar generates noticeable $L_Z-\langle V_R\rangle$ wave signals (extrema points for the phase shift measurement) only at the 2:1 and 1:1 OLR of the bar, which could make the number of measurable extrema points too limited to ascertain a clear $\Delta L_Z$ variation trend \citep[See Figure 11 of][]{chiba_etal_21}. On the other hand, generating more extrema points in the $L_Z-\langle V_R\rangle$ wave requires higher-order Fourier modes from the bar potential \citep{monari_etal_19} or a decrease in the pattern speed of the bar \citep{chiba_etal_21}, which is beyond the scope of this work. We exclude stars with an initial radius smaller than 1.5 kpc to save integration time. This will have little impact on the simulation results since the number of discarded particles capable of entering the investigated $L_Z$ range is negligible.

With the above setup, we divide each simulation snapshot into eight equally spaced azimuthal ranges and extract extrema points from the ``cold" and ``hot" waves. We focus on the $L_Z-\langle V_R\rangle$ wave pattern in the range of $L_Z \in [600,3000] \thinspace \mathrm{km \thinspace s^{-1}\thinspace kpc}$, divided into 165 equally spaced bins. The Gaussian kernel size used to smooth the $L_Z-\langle V_R\rangle$ wave for $\Delta L_Z$ measurement is $\sigma=4$. We compare the $\Delta L_Z$ variation trends in different azimuths to conclude a universal variation pattern. If there is none, we try to unveil the $\Delta L_Z$ variation feature existing in most azimuthal ranges, which may also be valuable for discriminative purposes. Due to the subtle and discrete nature of the measurable extrema points, conclusions drawn from the analysis on the $\Delta L_Z$ variation trend are reliable in the qualitative sense, but any quantitative conclusion should be treated with caution. Providing a quantitative match of the $\Delta L_Z$ variation curve between simulation results and observation data is difficult since the phase shift $\Delta L_Z$ is influenced by measurement settings, distribution function, lack of self-gravity, and other factors.

\subsection{External Satellite Perturbation} 
\subsubsection{Simulation Results For External Perturbation}  \label{subsec:ext}
\begin{figure*}
    \centering
    \includegraphics[width=\textwidth]{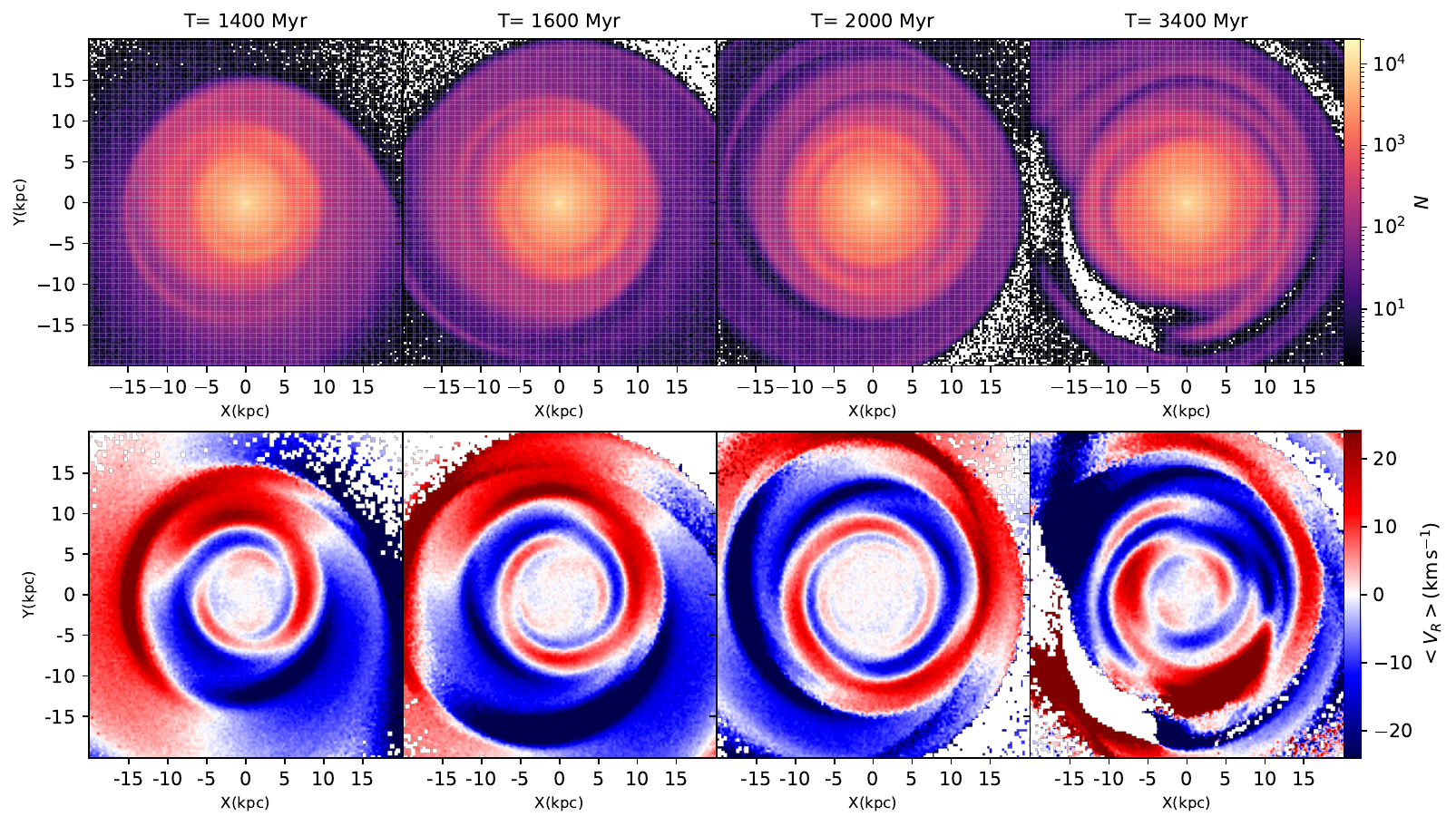}
    \caption{Temporal evolution of the modeled Galactic disk after the external satellite perturbation. The upper and lower rows are number density and the mean radial velocity $\langle V_R \rangle$ map in the $\mathrm{X-Y}$ plane. The evolution times of the four snapshots are 1400, 1600, 2000, and 3400 Myr from left to right. The first and second pericenter passages are at around 1 and 3.1 Gyr, respectively. As phase mixing proceeds, the spiral arms become more tightly wound.}
    \label{fig:ext_evol}
\end{figure*}

\begin{figure*}
    \centering
    \includegraphics[width=\textwidth]{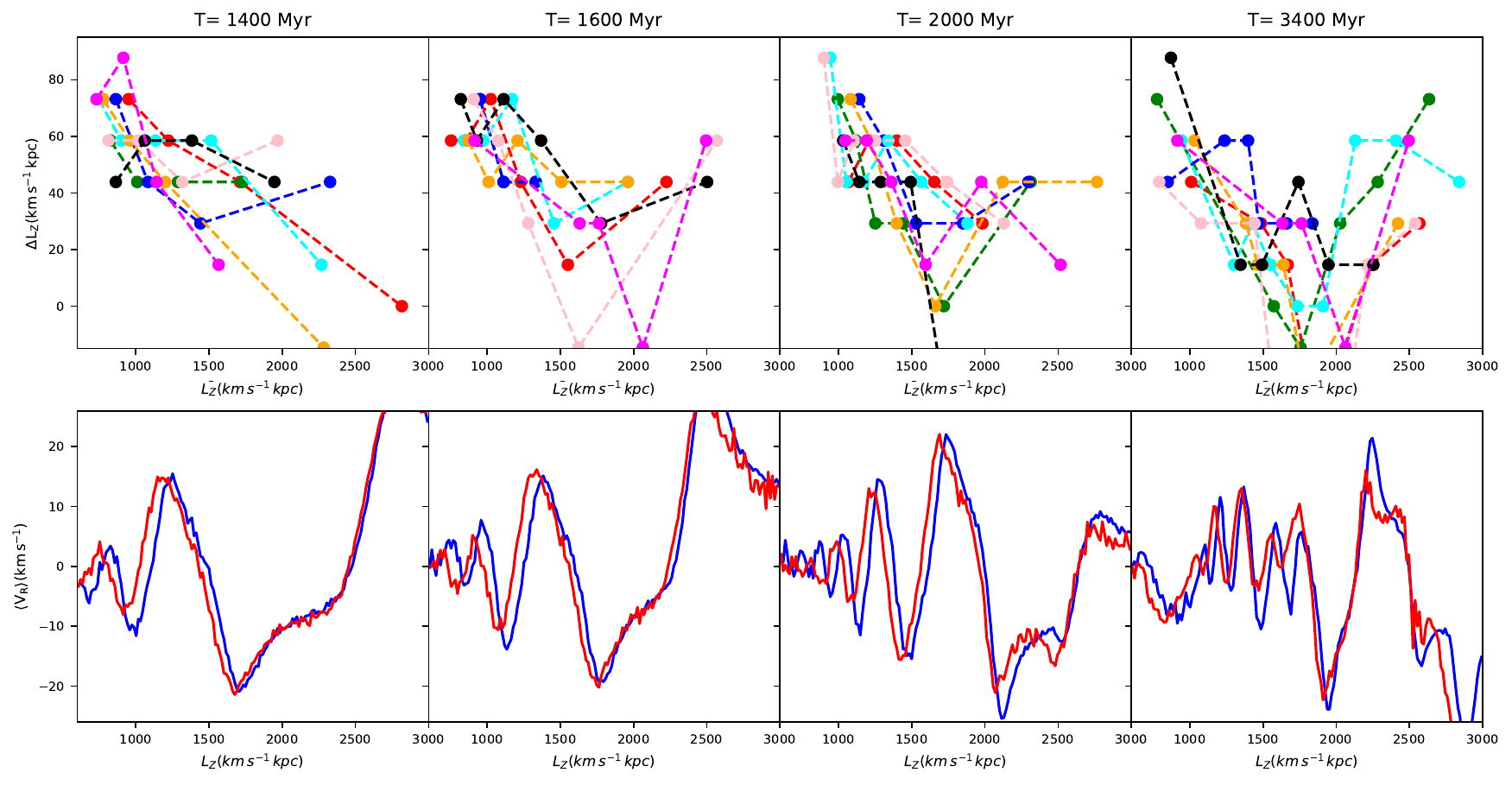}
    \caption{Phase shift amplitude $(\Delta L_Z)$ variation trend versus $L_Z$ after the satellite pericenter passage. The plots in the upper and lower panels are from the simulation data at 1400, 1600, 2000, and 3400 Myr. The dashed lines in the upper row illustrate the $\Delta L_Z-\bar{L_Z}$ plot in eight different azimuthal ranges. In the lower row, the red and blue solid lines display the $L_Z-\langle V_R\rangle$ wave signal for the dynamically hot ($J_Z\mathrm{>23 \thinspace km \thinspace s^{-1}\thinspace kpc}$) and cold ($J_Z\mathrm{<8 \thinspace km \thinspace s^{-1}\thinspace kpc}$) subpopulations. Within the lower $L_Z$ range, we see a monotonic decreasing trend of $\Delta L_Z$ with increasing $L_Z$ in most azimuthal ranges for all snapshots, consistent with the prediction of our toy model developed in Section \ref{subsec: model}.}
    \label{fig:ext_dlz_lz}
\end{figure*}

Figure \ref{fig:ext_evol} illustrates the evolution of the Milky Way-like disk after the satellite pericenter passage. Tidally induced spiral wraps are visible in the $X-Y$ projection of surface density and mean radial velocity $\langle V_R \rangle$. As time progresses, the spirals become more tightly wound. $\langle V_R \rangle$ of the innermost region is close to zero due to swifter phase mixing, which expands in spatial extent with time. In the meantime, the characteristic wavelength of the $L_Z-\langle V_R \rangle$ wave shown in Figure \ref{fig:ext_dlz_lz} increases with $L_Z$, consistent with our toy model in Section \ref{subsec: model} and the model of \cite{ramos_etal_22}. After the second pericenter passage, the spiral pattern exhibits more complex morphology, as shown in the rightmost column of Figure \ref{fig:ext_evol}. We do not show snapshots taken immediately after the satellite's pericenter passage because the number of measurable extrema points is too few to reveal a clear trend.

Two trends are visible in the $\bar{L_Z}-\Delta L_Z$ plot of Figure \ref{fig:ext_dlz_lz}. At a fixed time, $\Delta L_Z$ decreases as $L_Z$ (or $\Omega_R$) increases (or decreases) but rarely becomes negative in the lower $L_Z$ range. $\Delta L_Z$ at the higher $L_Z$ end is generally larger but does not exhibit a clear trend due to the lack of measurable extrema points. Besides, at a fixed $L_Z$ range, the phase shift amplitude $\Delta L_Z$ displays negligible variation with time. Our results unveil a novel formation channel for the phase shift in addition to the orbital resonance suggested by \cite{friske_19}. 

As the lower right panel of Figure \ref{fig:ext_dlz_lz} demonstrates, the $L_Z-\langle V_R \rangle$ wave signal exhibits less resemblance to the sinusoidal wave after the second pericenter passage. However, the decreasing $\Delta L_Z$ trend persists in the lower $L_Z$ range. The temporal variation of $\Delta L_Z$ is also negligible. Therefore, the same $\Delta L_Z$ variation trends described above persist as long as the external satellite remains the dominant perturber, regardless of the wave shape or the number of pericenter passages. 

The minor variation of $\Delta L_Z$ with time has important implications on two fronts. Firstly, it renders the utilization of $\Delta L_Z$ variation pattern to date perturbation infeasible. Fourier analysis technique proposed by \cite{ramos_etal_22} is more suitable for this task. On the other hand, this constancy suggests that this unique feature can aid in ascertaining the wave's origin regardless of the evolutionary stage.

\subsubsection{Toy model of radial phase mixing from External Perturbation} \label{subsec: model}
\begin{figure*}[ht]
    \centering
    \includegraphics[scale=0.4,width=\textwidth]{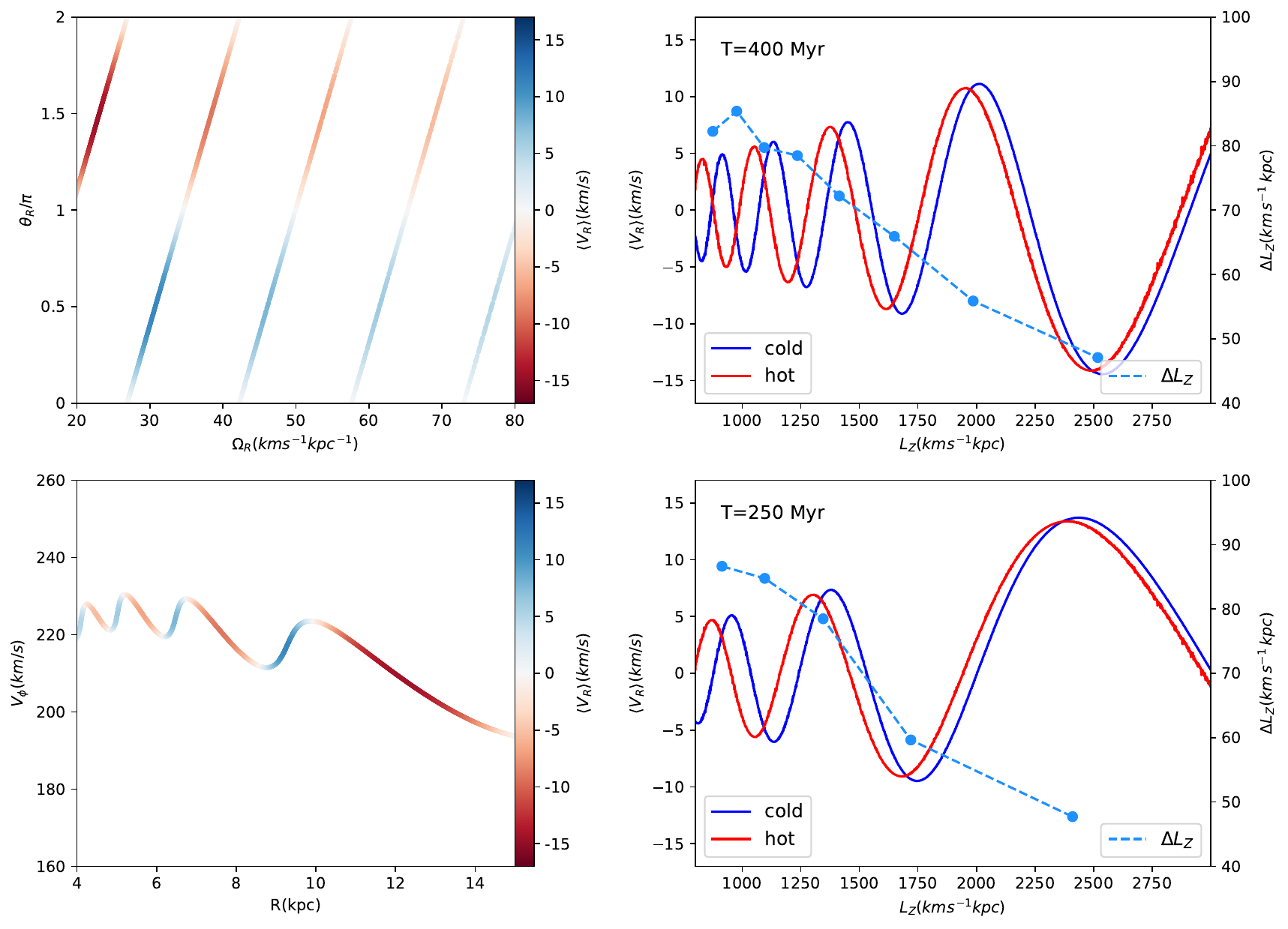}
    \caption{Illustration of the toy model of radial phase mixing presented in Section \ref{subsec: model}. The left panel maps the distribution of particles in the $\Omega_R-\theta_R$ and the $R-V_\phi$ space, both color coded by $\langle V_R\rangle$ at the time of 400 Myr. The right panel illustrates the $L_Z-\langle V_R\rangle$ wave signal of the dynamically cold ($J_Z\mathrm{<8\,km\,s^{-1}\,kpc}$, blue) and hot ($J_Z\mathrm{>23\,km\,s^{-1}\,kpc}$, red) populations at the time of 250 and 400 Myr. As phase mixing proceeds, the characteristic wavelength of the $L_Z-\langle V_R\rangle$ decreases. The wave of the dynamically hot population displays a systematic phase shift toward lower $L_Z$ compared to the cold one. The phase shift amplitude ($\Delta L_Z$, lighter blue points corresponding to the right axis) increases with decreasing $L_Z$, as predicted by our toy model. }
    \label{fig: model}
\end{figure*}

In this section, we introduce a toy model of radial phase mixing to elucidate the formation of phase space substructures following external perturbation. Additionally, we employ this model to qualitatively interpret the dependence of the $L_Z-\langle V_R\rangle$ wave on dynamical hotness. We represent each $L_Z$ bin with a single particle orbit, and the phase space coordinates of different orbits (corresponding to different $L_Z$ bins) are stitched together to depict regions of relatively high phase space density. The radial velocity $V_R$ of the orbit represents the mean radial velocity $\langle V_R \rangle$ of its corresponding $L_Z$ range.

To give the toy model predictive power in a realistic sense, we utilize the same initial condition as the test particle simulations in Section \ref{subsec:tps} and define the dynamically cold and hot sub sample in the same manner. We also calculate the mean values of radial frequency $\Omega_R$ for each $L_Z$ bin using \texttt{agama} \citep{agama_19}. The $\theta_R$ value is given by the equation
\begin{equation}
    \theta_R=\Omega_R\times T+\theta_{R,0}
    \label{eq:or_wr}
\end{equation}
where $T$ is the perturbation time. For simplicity, we assume both subpopulations follow the same distribution in the $\Omega_R-\theta_R$ plane. While this assumption is not strictly correct, it has a negligible impact on the key features of our prediction. By requiring all particles to start at the same radial phase $(\theta_{R,0}=0)$, we implicitly assume that all stars receive the impulsive perturbation from a distant perturber. Then we utilize epicycle approximation to compute the phase space coordinates for the particles using the following equations: 
\begin{equation}
    V_R=X_R\cos{\theta_R},R=R_g+X\sin{\theta_R}/\Omega_R
    \label{eq:vr_r}
\end{equation}
where $X_R$ represents the radial oscillation amplitude. \citep[See][for detailed derivation]{binney_08}. Upon adopting this equation, we also implicitly assume the particle orbits remain regular after the satellite perturbation. Then we map their distribution in the $L_Z-\langle V_R \rangle$ and the $R-V_\phi$ space to understand the structural correlation between them.

The numerical prescription is as follows. We choose particles with angular momentum $L_Z \in [800,3000]\,\mathrm{km\,s^{-1}\,kpc}$. The corresponding guiding radius $R_g$ is inferred from the rotation curve of the adopted Milky Way potential model \texttt{MWPotential2014} \citep{galpy_15}. The radial oscillation amplitude $X_R$ is in the form of $X_R=A\times R_g$, following \cite{struck_etal_11}. Changing the form of the dependence of $X_R$ on guiding radius $R_g$ has a negligible effect on the main results. The value of $A$ is set at 1.2 to ensure the $\langle V_R \rangle$ amplitude is at the same order of magnitude for both the observations and simulation results.

As the upper-left panel of Figure \ref{fig: model} shows, a series of parallel lines, each with the same slope set by the perturbation time, emerges in the $\Omega_R-\theta_R$ plane. In the bottom-left panel, ridge lines color coded by $V_R$ form connected V-shaped streaks. The adjacent branches of the V-shaped streaks display opposite signs of $\langle V_R\rangle$. Ridge lines reach a turning point when $\langle V_R\rangle$ changes sign and $\theta_R$ reaches 0 ($2\pi$) or $\pi$. These two features are qualitatively consistent with the $N$-body simulation of the interaction between an Sgr-like satellite and a Milky Way-like galaxy, as shown in Fig.18 of \cite{ramos_etal_22}. 

The $L_Z-\langle V_R\rangle$ wave also emerges, where $\langle V_R\rangle=0$ corresponds to a turning point of a V-shaped streak in the $R-V_\phi$ plane. The spacing between the adjacent zero radial velocity points becomes wider at higher $L_Z$ because of the shallower slope of the $L_Z-\Omega_R$ curve. The phase shift between waves of different dynamical hotness ($J_Z$) arises from differences in mean $\Omega_R$. Since we assume both populations follow the same $\Omega_R-\theta_R$ line series, this frequency difference leads to the phase difference in $\theta_R$. This phase difference increases toward the lower $L_Z$ range, where the difference in dynamical hotness (or mean $\Omega_R$) increases. As shown in the right panel of Figure \ref{fig: model}, the phase difference in $\theta_R$ increases, resulting in a shift in extrema location, with its amplitude ($\Delta L_Z$ as defined in Section \ref{subsec:data}) increasing with decreasing $L_Z$. As the right column of Figure \ref{fig: model} demonstrates, the slopes of the $\Omega_R-\theta_R$ line series become steeper with time, which causes the characteristic wavelength (or frequency) of the $L_Z-\langle V_R\rangle$ wave to decrease (or increase), which is qualitatively consistent with the simulation results shown in the previous subsection. \cite{ramos_etal_22} utilized this property to date the perturbation. From Equation \ref{eq:or_wr}, the phase difference in $\theta_R$ increases in the meantime. However, the characteristic wavelength of the $L_Z-\langle V_R\rangle$ wave also decreases, which cancels out the effect of increasing phase difference in $\theta_R$ and maintains a roughly constant phase shift amplitude. Since this phase shift accumulation with $L_Z$ is due to the difference in $\Omega_R$, its maximal value occurs in the low $L_Z$ end, rather than at the high $L_Z$ end where the external satellite perturbation is strongest.

Compared with Figure \ref{fig:ext_dlz_lz}, this toy model shows good agreement in terms of both the amplitude and variation trend of $\Delta L_Z$ in the lower $L_Z$ range, which verifies the key physical assumption of our toy model. The discrepancy at the high $L_Z$ end may be related to the irregular morphology of the orbits strongly perturbed by the satellite, which invalidates the usage of epicycle approximation in our toy model. This suggests that our toy model could not be applied to the outermost part of the disk. The negligible temporal variation of the $\Delta L_Z$ amplitude can also be explained by the two competing effects of the phase mixing, as described above.

\subsection{Internal Perturbers: Bar and Transient Spiral} \label{subsec:int}
\begin{figure}
    \centering
    \includegraphics[scale=0.65]{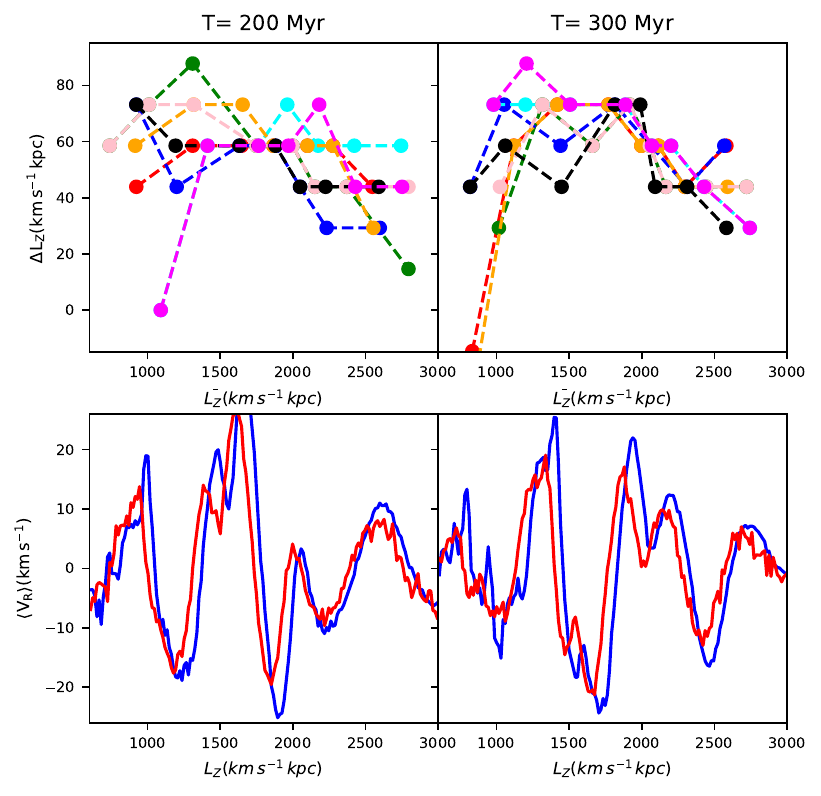}
    \caption{The temporal evolution of the modeled Galactic disk under the perturbation of two-armed transient spirals in the same manner as Figure \ref{fig:ext_dlz_lz}. The times are 200 and 300 Myr.}
    \label{fig:tsp_dlz_lz}
\end{figure}

In the case of the transient spirals, the phase shift amplitude $\Delta L_Z$ exhibits no universal trends with $L_Z$ among different azimuthal ranges. As Figure \ref{fig:tsp_dlz_lz} demonstrates, within the majority of azimuthal ranges, the phase shift amplitude $\Delta L_Z$ displays irregular oscillation with $L_Z$, with an amplitude between 30 and 80 $\mathrm{km\,s^{-1}\,kpc}$. We cannot observe any similar $\Delta L_Z$ variation pattern shared by both snapshots, either. Interestingly, the decreasing trend with $L_Z$ resembling the trend produced by external satellite perturbation exists for very few azimuthal ranges of the two snapshots. In such cases, the slope of the $\Delta L_Z$ curve is shallower than those generated from the satellite impact,  which still allows us to distinguish from the $\Delta L_Z$ variation trend of external perturbation. 

\begin{figure*}
    \centering
    \includegraphics[width=\textwidth]{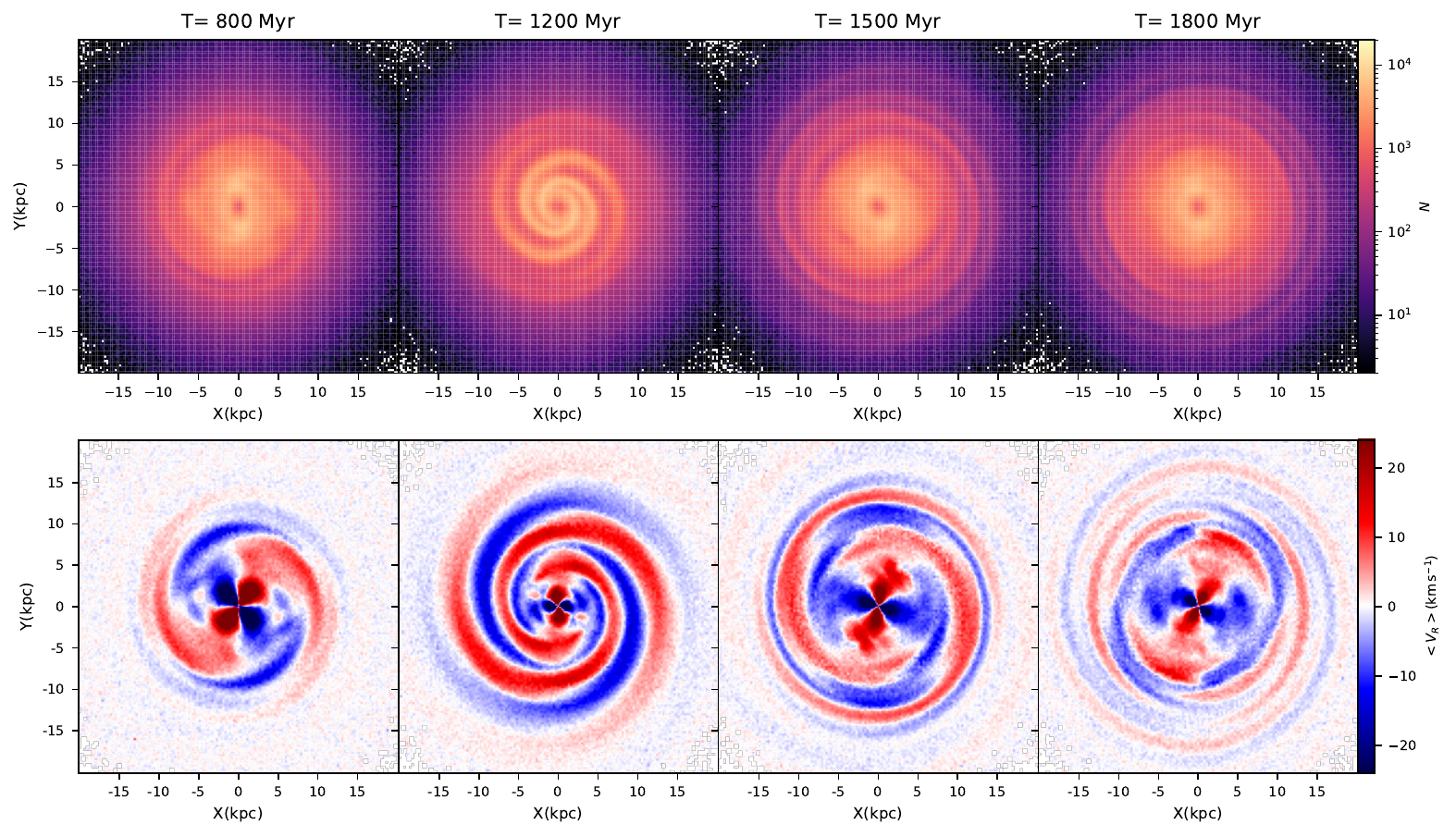}
    \caption{The temporal evolution of the modeled Galactic disk under the joint perturbation of transient spirals and a steadily rotating bar demonstrated in the same manner as Figure \ref{fig:ext_evol}. The transient spiral potential reaches its maximal amplitude at 1200 Myr.}
    \label{fig:int_evol}
\end{figure*}

The disk evolution shown in Figure \ref{fig:int_evol} is when the perturbations of the bar and transient spirals are combined. The central depression in density is the result of excluding the innermost particles from orbit integration. Before the transient spiral starts growing ($T\mathrm{< 1000 \, Myr}$), the disk dynamics are mainly influenced by the bar, as shown by the butterfly pattern in the $\langle V_R \rangle$ map. When the spiral potential reaches maximal amplitude at 1200 Myr, the two-armed spiral pattern becomes most prominent in both the density and $\langle V_R \rangle$ maps. The amplitude of $\langle V_R \rangle$ decreases as the spirals wrap up and decay. The bar resumes domination as the time approaches the end of our simulation. 

\begin{figure*}
    \centering
    \includegraphics[width=\textwidth]{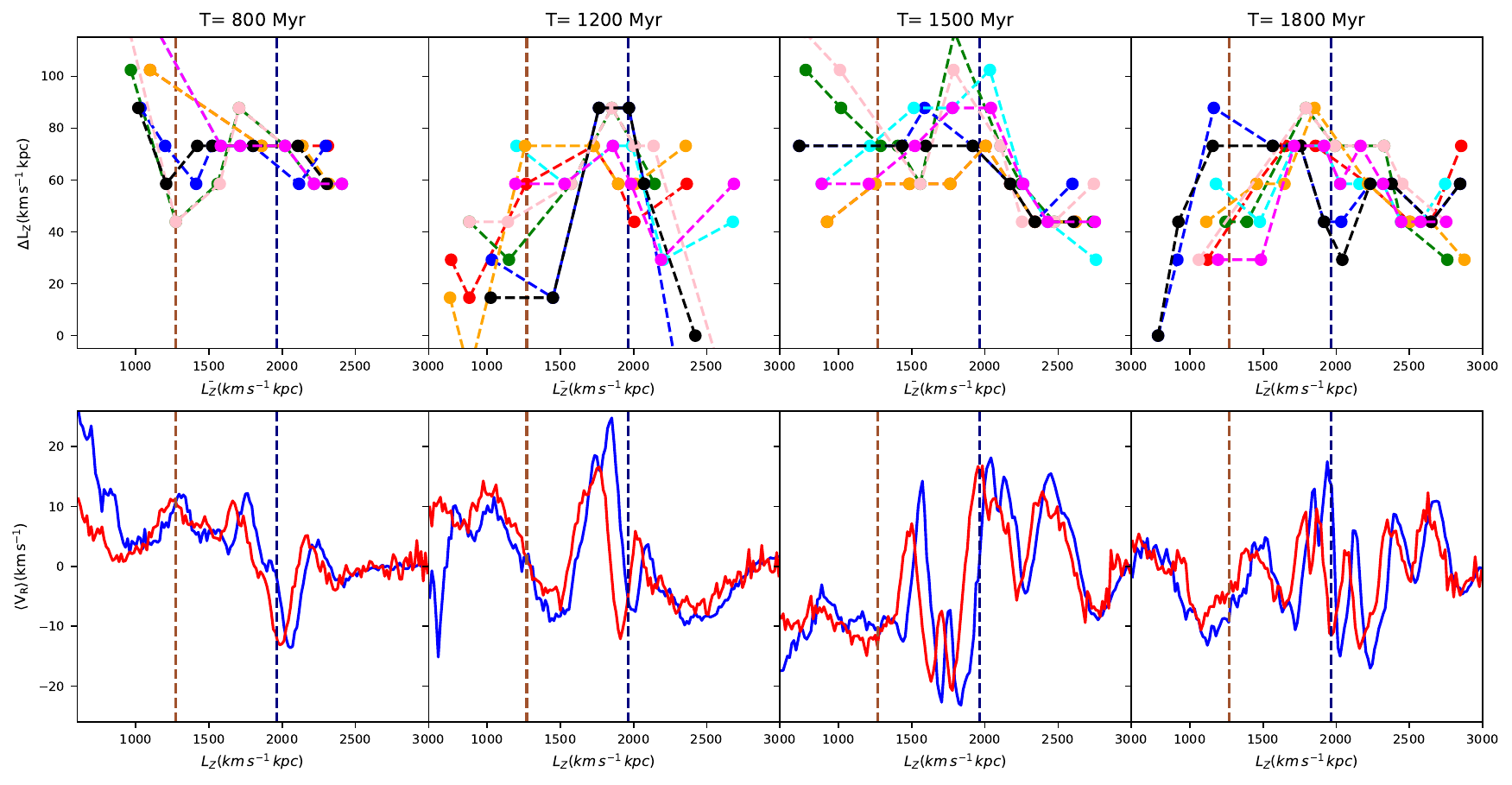}
    \caption{The phase shift $(\Delta L_Z)$ variation trend is shown in the same manner as Figure \ref{fig:ext_dlz_lz}. The four snapshots are from simulation data at 800 (represents the bar-only case for comparison), 1200, 1500, and 1800 Myr. The vertical dashed lines mark the locations of corotation resonance (brown) and the 2:1 OLR (blue) of the bar.}
    \label{fig:btsp_dlz_lz}
\end{figure*}

As shown in the leftmost column of Figure \ref{fig:btsp_dlz_lz}, when only the bar potential is active, the $\Delta L_Z$ variation trends in different azimuthal ranges do not share common features. The $\Delta L_Z$ amplitude exhibits smaller dispersion than the transient spiral case and increases to a larger value at the $L_Z$ range below the corotation resonance. However, the wave shape consistency between the two subpopulations becomes worse within the lower $L_Z$ range, making this feature susceptible to systematic error.

After the spiral potential becomes active, although the overall $\Delta L_Z$ pattern illustrated in the three right columns of Figure \ref{fig:btsp_dlz_lz} varies drastically with time, a common feature in the phase shift variation pattern emerges across most of the azimuthal ranges: a characteristic peak at $L_Z\sim 2000\, \mathrm{km\,s^{-1}\,kpc}$ with variable prominence. Interestingly, the location of the $\Delta L_Z$ peak coincides with the location of the 2:1 OLR of the bar (marked by a vertical blue dashed line) in most azimuthal ranges. In a minority of the azimuthal ranges, the $\Delta L_Z$ peak is less prominent or absent, which makes the $\Delta L_Z$ variation trend qualitatively similar to those produced by the transient spirals. Despite the caveats mentioned above, the sufficiently prominent $\Delta L_Z$ peak could provide the most likely range of the 2:1 OLR to constrain the bar pattern speed $\Omega_b$.

Due to the fundamental difference between internal and external perturbations, our toy model of radial phase mixing cannot account for the $\Delta L_Z$ variation trends generated by the bar and spirals, which may be caused by the dependence of resonance lines on dynamical hotness ($J_Z$). As illustrated in Figure 10 of \cite{trick_etal_21}, the location of bar resonance (the so-called ``ARL", axisymmetric resonance line) shifts toward lower $L_Z$ for stars of higher $J_Z$. In the region of the bar's 2:1 OLR, the majority of stars are forced toward higher $L_Z$, and the changes of $L_Z$ ($\delta L_Z[L_{Z,0}]$) are smaller than corotation resonance. Within corotation resonance, radial ($L_Z$) migration can proceed in both directions, which could partially cancel out the effect of the ARL shift. This is not the case for Lindblad resonances. Therefore, the bar's 2:1 OLR can generate a distinct $\Delta L_Z$ peak in the presence of the spirals. No prominent $\Delta L_Z$ peaks emerge under transient spiral perturbation since most stars approximately corotate with the spirals and are influenced by corotation resonance.  However, the explanation above cannot account for the absence of the $\Delta L_Z$ peak feature in the bar-only simulation. More comprehensive theoretical studies are needed to fully comprehend the complex interplay between spiral and bar perturbations in the future.

To briefly summarize, the $L_Z-\langle V_R \rangle$ wave pattern displays different dependence on dynamical hotness depending on the nature of dominant perturbations. Utilizing the amplitude of phase shift ($\Delta L_Z$) between the wave patterns of different dynamical hotness, we could quantify this dependence and associate its variation trend versus $L_Z$ with particular perturbation types. The external satellite generates a general trend of decreasing $\Delta L_Z$ in the lower $L_Z$ range in almost all azimuths during the whole time. Such universality does not exist in the case of the transient spirals or the bar. Under the joint perturbation of the bar and the transient spiral, $\Delta L_Z$ displays a characteristic peak around 2:1 OLR, which qualitatively reproduces the observed $\Delta L_Z$ feature. Controlled tests demonstrate that the internal perturbation (the combination of the bar and the spirals) is more favorable than the external one to reproduce the observed phase shift variation pattern. 

\section{Discussion} \label{sec:discu}
\subsection{Limitations of Our Models}
Despite capturing the essential physical processes of radial phase mixing, our toy model presented in Section \ref{subsec: model} is not exempt from limitations. Firstly, it outlines the backbone of the phase space substructure but falls short of providing a detailed phase space density distribution. As the system becomes fully phase mixed in the lower $L_Z$ range, the $\langle V_R \rangle$ amplitude should be close to zero, which is not seen in the toy model. Neglecting the azimuthal dependence of the external perturbation, our toy model does not reflect the azimuthal phase-mixing process as the model of \cite{ramos_etal_22} does. Employing the epicycle approximation, our model can only account for the $\Omega_R$ difference between cold and hot orbits, prohibiting us from explaining the larger $\Delta L_Z$ at the high $L_Z$ end.

It is also imperative to be cautious while interpreting the $\Delta L_Z$ variation trends generated by test particle simulations, given the absence of self-gravity effects. Future investigations employing tailored $N$-body simulations \citep[e.g.,][]{jbh_21,hunt_etal_21} are required to validate our main results.

We should note that the $\Delta L_Z$ variation trend can only differentiate different perturbations 
in the probabilistic sense because not all azimuthal ranges exhibit the common $\Delta L_Z$ variation features unique to specific perturbations. Under the satellite impact, the decreasing $\Delta L_Z$ trend predicted by our toy model is less pronounced in some of the azimuthal ranges. In the case of bar and spiral arms, the 2:1 OLR of the bar does not generate the characteristic $\Delta L_Z$ peak in all azimuthal ranges. Therefore, broader azimuthal coverage of the Galactic disk in future surveys is essential for robust diagnostics of the perturbation history.

\subsection{Comparisons with Previous Works}

\cite{friske_19} measured the phase shift by fitting the observed data with the sinusoidal function and comparing the best-fit phase angles with each other. The unit of their phase shift is the degree, not the angular momentum used ($\mathrm{km\,s^{-1}\,kpc}$) in our work. While their approach is advantageous in utilizing complete wave information, it is susceptible to systematic errors due to its model-dependent nature. During fitting, they set the wavelength of the sine function to fixed values, essentially making their approach akin to peak finding in principle. Our method does not assume any waveform, but the results of peak finding could vary with the width of $L_Z$ bins. Both approaches have the issue of defining the phase shift when the wave shapes deviate significantly from the sinusoidal shape. New mathematical tools that can robustly measure the phase shift in such nonsinusoidal waves are indispensable for future research.

\subsection{Implications for the Phase Shift}
The comparison between the observational data and our test particle simulation results suggests that the $L_Z-\langle V_R\rangle$ wave is more likely to originate from internal perturbations rather than external satellites. In the case of external perturbation, extrema points with the most prominent $\Delta L_Z$ are located at both ends of the investigated $L_Z$ range, which disagrees with the key observational feature. In contrast, the characteristic $\Delta L_Z$ peak generated by the 2:1 OLR of the bar persists regardless of the evolution phase of the transient spiral arms. Therefore, if the observed phase shift variation trend does represent the trends within most of the azimuthal ranges, the combination of the bar and the transient spiral is more likely to be the dominant driving force. It is important to note that this conclusion is not definitive, as other scenarios, such as a decelerating bar \citep{chiba_etal_21} have not been explored in this work. Therefore, the $\Delta L_Z$ variation trend is not a sufficient but necessary criterion to discriminate different perturbing mechanisms. Nevertheless, our study provides a novel approach to differentiating the roles of internal and external perturbations in sculpting the observed phase space substructures.

According to our simulation results, the combination of the bar and the transient spiral arms is more likely to be the dominant perturbation, where the peak of the phase shift amplitude $\Delta L_Z$ is associated with 2:1 OLR. This $\Delta L_Z$ diagnostic could pave a new way to constrain the pattern speed of the Galactic bar. Utilizing test particle simulations with bar as the sole perturber, \cite{trick_etal_21} \citep[also in][] {muhlbauer_03} proposed that the $\langle V_R \rangle$ map in the $L_Z-J_R$ action space displaying the sign change is the signature of 2:1 OLR. However, owing to the interference of the transient spirals, the sign change in the $L_Z-\langle V_R \rangle$ wave pattern could also occur in $L_Z$ ranges other than specific types of resonances, as demonstrated by \cite{hunt_etal_19} in the $R-V_\phi$ projection. Our work could help break this degeneracy since it provides a viable way to detect the bar resonance signature in the presence of spirals. Among the four candidates of the 2:1 OLR of the bar proposed by \cite{trick_etal_21}, the ``Hat" and ``Sirius" moving groups are the two closest to the $\Delta L_Z$ peak of the observational data, i.e. the 2:1 OLR of the bar. This gives a crude pattern speed estimate of $\sim 33-41$ $\mathrm{km \, s^{-1} \,kpc^{-1}}$, which favors the long/slow bar model. The above constraint is consistent with previous works using the solar neighborhood kinematics to indirectly infer the bar pattern speed \citep{binney_20,chiba_21_2,trick_22}. The uncertainty of our estimation is greater than conventional methods because we can only give the most probable $L_Z$ range of the 2:1 OLR location from discrete extrema points. 

\cite{wang_etal_20} classified the $R-V_{\phi}$ ridges into two types depending on whether they exhibit significant variation with stellar age in the $\langle V_R\rangle$ map. Coincidentally, the ridge displaying the most prominent variation with age roughly follows the constant $L_Z$ line at $\mathrm{2180 \thinspace km \thinspace s^{-1} \thinspace kpc}$ where $\Delta L_Z$ exhibits a peak feature. Given the tendency of older stars to be dynamically hotter, this prominent variation could be interpreted through the 2:1 OLR of the Galactic bar, from the dynamical perspective. 

\section{Conclusion} \label{sec:conclu}
Decoding the dynamical evolution history of the Milky Way is challenging due to the degeneracy caused by various perturbations. Our work attempts to shed new light on this task by differentiating the formation mechanism of the recently found phase space substructures within the Galactic disk. The $L_Z-\langle V_R \rangle$ wave pattern is the one-dimensional deprojection of $R-V_\phi$ ridges and an easier target for comparative analysis. Leveraging the Gaia DR3 data and controlled numerical experiments using test particle simulations, we find that the variation trend of phase shift with $L_Z$ between the $L_Z-\langle V_R\rangle$ waves of different dynamical hotness could be an indicator of the dominant perturbations. With a simple toy model, our work also helps better comprehend the role of internal and external perturbers in shaping phase space substructures within the Galactic disk.

Our key findings are summarized as follows:

1. Analysis of the Gaia DR3 data reveals the $L_Z-\langle V_R\rangle$ wave of the dynamically hotter (larger vertical action $J_Z$) population systematically shifts toward lower $L_Z$. Defining this phase shift as the difference in extrema location between the wave pattern of the dynamically cold and hot populations, its amplitude $\Delta L_Z$ exhibits a prominent peak feature at $L_Z\sim $ \textbf{2100}$\, \mathrm{km\,s^{-1}\,kpc}$.

2. The external satellite perturbation produces a decreasing trend with increasing $L_Z$ for $\Delta L_Z$ in the lower $L_Z$ range, which validates the theoretical prediction given by our toy model of radial phase mixing. This trend arises from the increase of radial frequency $\Omega_R$ difference between sub-populations of different dynamical hotness with decreasing $L_Z$. The persistence of this trend regardless of the number of pericenter passages the satellite experienced makes it a unique signature of external perturbation.

3. The transient spirals or the bar alone do not produce a universal $\Delta L_Z$ variation trend with $L_Z$ for most azimuthal ranges. Combined with a steadily rotating bar, a characteristic $\Delta L_Z$ peak emerges at the 2:1 OLR of the bar in most azimuthal ranges, which qualitatively resembles the observational feature.

4. Comparisons between observation data and test particle simulation results suggest an internal cause for the observed $L_Z-\langle V_R\rangle$ wave, i.e. likely from the bar and spiral arms. If that is the case, the $\Delta L_Z$ peak marks the location of the 2:1 OLR of the bar, which supports a long/slow bar model with pattern speed between 33 and 41 $\mathrm{km \thinspace s^{-1} \thinspace kpc^{-1}}$.

In conclusion, we demonstrate that examining the phase shift behaviors between dynamically hot and cold stellar populations holds the potential to differentiate between different types of perturbations. Additionally, by pinpointing a likely internal origin for the phase shift variation trends, our work provides constraints on the relative importance of internal and external processes in the recent dynamical evolution history of the Galactic disk.

While our study represents significant progress, limitations exist that motivate several promising avenues for future investigations. For instance, tailored $N$-body simulations are required to investigate the self-gravity effects absent in test particle models. Reliable mathematical techniques must be developed for robustly measuring phase shifts, especially for nonsinusoidal wave shapes. Upcoming work can also examine the $\Delta L_Z$ variation patterns when combining different perturbations in test particle simulations.  

Future spectroscopic surveys such as 4MOST and Milky Way Mapper covering a wider range in spatial extent will be capable of uncovering the azimuthal variation of phase shift variation pattern, which could help constrain possible perturbation scenarios.
\section{Acknowledgments}
We thank the referee for helpful suggestions that improved the quality and structure of the paper. This work is supported by the National Natural Science Foundation of China under grant No.12122301 and 12233001, by a Shanghai Natural Science Research Grant (21ZR1430600), by the ``111'' project of the Ministry of Education under grant No. B20019, and by the China Manned Space Project with No. CMS-CSST-2021-B03. ZYL thanks the sponsorship from the Yangyang Development Fund. RS gratefully acknowledges the generous funding of a Royal Society University Research Fellowship. TA acknowledges the grant RYC2018-025968-I funded by MCIN/AEI/10.13039/501100011033 and by ``ESF Investing in your future’’, the grants PID2021-125451NA-I00 and CNS2022-135232 funded by MICIU/AEI/10.13039/501100011033 and by ``ERDF A way of making Europe’’, by the ``European Union'' and by the ``European Union Next Generation EU/PRTR'' and the Institute of Cosmos Sciences University of Barcelona (ICCUB, Unidad de Excelencia ’Mar\'{\i}a de Maeztu’) through grant CEX2019-000918-M.

This work has made use of data from the European Space Agency (ESA) mission Gaia (\url{https://www.cosmos.esa.int/gaia}), processed by the Gaia Data Processing and Analysis Consortium (DPAC, \url{https://www.cosmos.esa.int/web/gaia/dpac/consortium}). Funding for the DPAC has been provided by national institutions, in particular the institutions participating in the Gaia Multilateral Agreement. This work made use of the Gravity Supercomputer at the Department of Astronomy, Shanghai Jiao Tong University.

%


\software{astropy \citep{2013A&A...558A..33A,2018AJ....156..123A,astropy_22}, agama \citep{agama_19}, matplotlib \citep{Hunter_07}, galpy \citep{galpy_15}
          }
\facilities{Gaia}



\bibliography{sample631}{}
\bibliographystyle{aasjournal}


\end{CJK}
\end{document}